\documentclass[twocolumn,amsmath,amssymb,showpacs,floats,prl,aps,unsortedaddress,superscriptaddress]{revtex4-1}

\usepackage{graphicx}
\usepackage{float}
\usepackage{hyperref}
\usepackage{natbib}
\usepackage{bm}
\usepackage{tabularx}
\usepackage{color}
\usepackage{pdfpages}

\newcolumntype{Z}{>{\centering\let\newline\\\arraybackslash\hspace{0pt}}X}

\makeatletter
\AtBeginDocument{\let\LS@rot\@undefined}
\makeatother

\begin{document}
\title{$\mathbf{J}$-freezing and Hund's rules in spin-orbit-coupled multiorbital Hubbard models}
\author{Aaram J. Kim}
\email{aaram@itp.uni-frankfurt.de}
\affiliation{Institut f\"ur Theoretische Physik, Goethe-Universit\"at
  Frankfurt, Max-von-Laue-Str. 1, 60438 Frankfurt am Main, Germany}

\author{Harald O. Jeschke} 
\affiliation{Institut f\"ur Theoretische Physik, Goethe-Universit\"at
  Frankfurt, Max-von-Laue-Str. 1, 60438 Frankfurt am Main, Germany}

\author{Philipp Werner} 
\affiliation{Department of Physics, University of Fribourg, Chemin du
  Mus\'ee 3, 1700 Fribourg, Switzerland}

\author{Roser Valent\'{\i}} 
\affiliation{Institut f\"ur Theoretische Physik, Goethe-Universit\"at
  Frankfurt, Max-von-Laue-Str. 1, 60438 Frankfurt am Main, Germany}

\begin{abstract}
  We investigate the phase diagram of the
  spin-orbit-coupled three orbital Hubbard model at arbitrary filling
  by means of dynamical mean-field theory combined with
  continuous-time quantum Monte Carlo.  
  We find that the spin-freezing crossover
  occurring in   the metallic phase of 
  the non-relativistic multiorbital Hubbard model can be
  generalized to a $\mathbf{J}$-freezing crossover, with
  $\mathbf{J}=\mathbf{L}+\mathbf{S}$, in the spin-orbit-coupled
  case. 
  In the $\mathbf{J}$-frozen regime 
  the correlated electrons exhibit a non-trivial flavor selectivity
and energy dependence.
  Furthermore, in the regions near  $n=2$ and $n=4$ the metallic
  states are qualitatively different from each other, which reflects the atomic Hund's third rule.
  Finally, we explore the appearance of  magnetic order from
  exciton condensation at $n=4$ and discuss the relevance of our results for real materials.
\end{abstract}


\maketitle

{\it Introduction.}
In $4d$ and $5d$ transition metal oxides the interplay and competition
between 
kinetic energy, spin-orbit coupling (SOC) and correlation effects
results in several interesting phenomena, such as spin-orbit assisted
Mott
transitions~\cite{Kim2009phase,Chaloupka:2010gi,comin20122,plumb2014alpha,Steve:2016,Lang:2016},
unconventional superconductivity~\cite{Meng:2014jo,Chaloupka:2016gx},
topological phases~\cite{WitczakKrempa:2014hz}, exciton
condensation~\cite{Khaliullin:2013du,Chaloupka:2016gx,Meetei:2015bu},
or exotic magnetic orders~\cite{Chen:2011bv,Chen:2010ec}.  Transition
metal oxides involving $4d$ and $5d$ electrons show diverse structures
like the Ruddlesden-Popper series~\cite{Kim2009phase,Meng:2014jo},
double perovskite,~\cite{Chen:2011bv,Chen:2010ec,Meetei:2013fd}
two-dimensional honeycomb
geometry~\cite{Singh2010,comin20122,plumb2014alpha,Coldea2016,Steve:2016,Lang:2016} or
pyrochlore lattices~\cite{Shinaoka:2015cg}. In an octahedral
environment, as in most of the 4$d$ and 5$d$ materials mentioned
above, the five $d$ orbitals are split into low energy $t_{2g}$ and
higher energy $e_g$ levels.  The SOC further splits the low energy
$t_{2g}$ levels into a so-called $j=1/2$ doublet and $j=3/2$
quadruplet.  The energy separation between the $j=1/2$ and $j=3/2$ bands
is proportional to the strength of the SOC.  Existing
 {\it ab-initio} density functional theory
calculations~\cite{Shinaoka:2015cg,Foyevtsova2013} suggest that in some
materials a multiorbital
description including both the $j=1/2$ and $j=3/2$ subbands should be
considered.

Most theoretical studies of 4$d$ and 5$d$ systems
have focused on material-specific models with
fixed electronic filling. Here we follow a different strategy and
explore the possible states that emerge from a multiband Hubbard
model with spin-orbit coupling at arbitrary filling. This allows us to
investigate unexplored regions in parameter space 
which may exhibit interesting phenomena.
Specifically, by performing a systematic analysis of the local
$\mathbf{J}$ moment susceptibility ($\mathbf{J}=\mathbf{L}+\mathbf{S}$)
as a function of Coulomb repulsion $U$, Hund's
coupling $J_{\rm H}$, spin-orbit coupling $\lambda$ and filling $n$,
we identify Mott-Hubbard insulating phases and complex metallic
states.  We find a $\mathbf{J}$-freezing crossover between 
 a Fermi liquid (FL) and a 
non-Fermi liquid (NFL) phase where the latter shows a 
 distinct flavor selectivity that originates from the SOC.
 In addition, we observe a strong asymmetry in the metallic phase between 
 filling $n=2$ and $n=4$ with properties reminiscent of the atomic Hund's third rule.  
Finally, we investigate doping effects on the excitonic magnetism at $n=4$.

\begin{figure*}[ht]
  \includegraphics[width=1.0\textwidth]{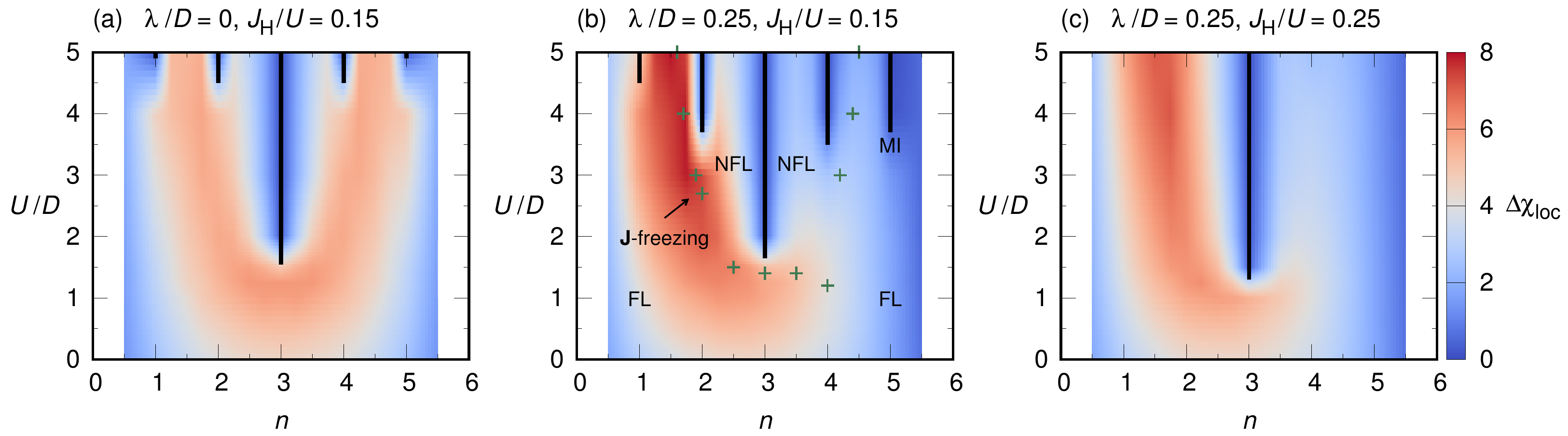}
  \caption{Dynamic contribution to the local susceptibility,
    $\Delta\chi_\text{loc}$ in the $(U/D,n)$ phase diagram for (a)
    $\lambda/D=0.0$, $J_{\rm H}/U=0.15$, (b) $\lambda/D=0.25$, $J_{\rm
      H}/U=0.15$, (c) $\lambda/D=0.25$, $J_{\rm H}/U=0.25$, and
    $T/D=0.03$. 
    Cross symbols mark the maximum values of
    $\Delta\chi_\text{loc}$ corresponding to the $\mathbf{J}$-freezing
    crossover points.  
    The parameter set for (b) is chosen by following the tight-binding parameters of Sr$_2$IrO$_4$~\cite{Watanabe:2010}.
    The reported values of the $\lambda$, $J_\text{H}$, and $U$ for various materials are summarized in the supplementary material.
    }
  \label{fig:PhaseDiagram}
\end{figure*}

{\it Method.}
We consider a three-orbital Hubbard model with spin-orbit coupling.
The model Hamiltonian consists of three terms,
\begin{equation}
  \mathcal{H} = \mathcal{H}_t + \mathcal{H}_{\lambda} + \mathcal{H}_{U}~,
  \label{eqn:model}
\end{equation}
where $\mathcal{H}_t$, $\mathcal{H}_{\lambda}$, and $\mathcal{H}_{U}$
denote the electron hopping, spin-orbit coupling, and local Coulomb
interaction terms, respectively.  
In order to discuss the underlying physics, 
relevant for a range of materials with different structures, 
we use a semi-circular density of states (DOS), 
$\rho^0(\omega) = (2/\pi D)\sqrt{1-(\omega/D)^2}$ for all orbitals.
The half-bandwidth $D$ is set to unity.
$\mathcal{H}_{\lambda}$ is
constructed by projecting the SOC term of $d$ orbitals onto the
$t_{2g}$ subspace, 
\begin{equation}
	\mathcal{H}_{\lambda} = \lambda \sum^{}_{\substack{\alpha\beta\\\sigma\sigma'}}c^{\dagger}_{i\alpha\sigma}\langle \alpha\sigma|\mathcal{P}_{t_{2g}}\mathbf{L}^{d}\mathcal{P}_{t_{2g}}\cdot \mathbf{S}|\beta\sigma'\rangle c^{}_{i\beta\sigma'}~,
\label{eqn:Hsoc}
\end{equation}
where $\mathcal{P}_{t_{2g}}$ is the projection operator.
$c^{}_{i\alpha\sigma}$ ($c^{\dagger}_{i\alpha\sigma}$) denotes the
annihilation (creation) operator of a spin $\sigma$ electron at site
$i$ and  orbital $\alpha$.  The angular momentum operator within the
$t_{2g}$ subspace can be represented by an effective $L=1$ angular
momentum operator with an extra minus sign~\cite{Chen:2011bv}.

The local Coulomb interaction Hamiltonian is written in Kanamori
form~\cite{Kanamori:1963} including the spin-flip and pair-hopping
terms as:
\begin{eqnarray}
  \mathcal{H}_{U} &=& U\sum^{}_{i,\alpha}n_{i\alpha\uparrow}n_{i\alpha\downarrow} + \sum^{}_{\substack{i,\alpha<\alpha' \\ \sigma\sigma'}}(U'-J_{\rm H}\delta_{\sigma\sigma'})n_{i\alpha\sigma}n_{i\alpha'\sigma'} \nonumber\\
  && -J_{\rm H}\sum^{}_{i,\alpha<\alpha'}(c^{\dagger}_{i\alpha\uparrow}c^{\dagger}_{i\alpha'\downarrow}c^{}_{i\alpha'\uparrow}c^{}_{i\alpha\downarrow} + h.c.)\nonumber\\
  && +J_{\rm H}\sum^{}_{i,\alpha<\alpha'}(c^{\dagger}_{i\alpha\uparrow}c^{\dagger}_{i\alpha\downarrow}c^{}_{i\alpha'\downarrow}c^{}_{i\alpha'\uparrow} + h.c.)~.
\end{eqnarray}
Here, $U$ is the on-site Coulomb interaction and $J_\text{H}$ denotes the Hund's coupling.
$U'$ is set to $U-2J_\text{H}$ to make the interaction rotationally invariant
in orbital space.

We employ the dynamical mean-field theory (DMFT)~\cite{Georges:1996zz}
to solve the model Hamiltonian Eq.~(\ref{eqn:model})
in a broad parameter space.  Since DMFT is a
nonperturbative technique within the local self-energy approximation, we can
access metallic and insulating phases on the same footing.
  In addition, the dynamical
fluctuations encoded in the DMFT solution contain valuable information
on the degree of moment correlations and the corresponding
susceptibility.  We will use the local $\mathbf{J}$ moment susceptibility as a
central quantity to investigate the phase diagram.

As an impurity solver, we adopt the continuous-time quantum Monte Carlo method
(CTQMC) in the hybridization expansion variant~\cite{Gull:2011jd,Werner:2006iz}.  For the single particle
basis of the CTQMC calculation, we choose the relativistic $j$
effective basis ($j=1/2$, $j=3/2$) which is an eigenbasis of the SOC
Hamiltonian.
It was previously reported that the $j$ effective basis reduces the
sign problem of the CTQMC simulation~\cite{Sato:2015hq}.  
For symmetry broken phases,
we consider the off-diagonal hybridization functions.

{\it Results.} 
A strong Coulomb interaction localizes electrons and can lead to the
formation of local moments.  The freezing of these local moments is
signaled by a slow decay, and eventual saturation, of the dynamical
correlation function $\langle J_z(\tau)J_z(0)\rangle$ on the imaginary-time axis. 
Hence, the local susceptibility, defined as
\begin{equation}
  \chi_\text{loc} = \int_{0}^{\beta}d\tau~\langle J_z(\tau)J_z(0)\rangle~,
  \label{eqn:chi}
\end{equation}
allows us to investigate the formation and freezing of local moments.
In addition, we define the dynamical contribution to the local
susceptibility by eliminating the long-term memory of the correlation
function from the original $\chi_\text{loc}$~\cite{Hoshino:2015gb}:
\begin{equation}
  \Delta\chi_\text{loc} = \int_{0}^{\beta}d\tau~\Big(\langle J_z(\tau)J_z(0)\rangle - \langle J_z(\beta/2)J_z(0)\rangle\Big)~.
  \label{eqn:Dchi}
\end{equation}
As the system evolves from an itinerant to a localized phase,
$\Delta\chi_\text{loc}$ exhibits a maximum in the intermediate Coulomb
interaction regime~(see Fig.~S1(c,d) in Ref.~\cite{Aaram:2016supp}); both, (i) the enhanced
correlations compared to the noninteracting limit and (ii) the larger
fluctuations compared to the localized limit lead to the maximum in
$\Delta\chi_\text{loc}$.  The location of the $\Delta\chi_\text{loc}$
maxima in the phase diagram can be viewed as the boundary of the local
moment regime and has been used to define the `spin-freezing crossover
line' in the non-spin-orbit coupled
system~\cite{Werner:2008fu,Hoshino:2015gb}.  However, since spin is
not a good quantum number in the spin-orbit-coupled system, we
introduce the total moment $\mathbf{J}=\mathbf{S}+\mathbf{L}$ to 
generalize the  `spin-freezing' 
 to a `$\mathbf{J}$-freezing' crossover.

In the following, we discuss the paramagnetic phase diagram of 
Eq.~(\ref{eqn:model}) obtained with DMFT(CTQMC) as a function of
 $U$,  $J_{\rm H}$,
 $\lambda$ and $n$. 
Figures~\ref{fig:PhaseDiagram} (a-c) show contour plots of
$\Delta\chi_\text{loc}$ in the interaction vs. filling plane for three
different parameter sets of $\lambda$ and $J_{\rm H}$.  Since SOC
breaks particle-hole symmetry, Fig.~\ref{fig:PhaseDiagram}~(b), (c)
are not symmetric about the half-filling axis, $n=3$. 
The Mott insulating phase (black lines in
Fig.~\ref{fig:PhaseDiagram}) which we identify as the region where the
spectral function vanishes at the Fermi-level and where
$\Delta\chi_\text{loc}$ is smallest, appears at each commensurate
filling. Nonetheless, compared to the system without SOC
(Fig.~\ref{fig:PhaseDiagram} (a)), the change of the critical
interaction strength $U_{\rm c}$ shows a complex behavior depending on
the filling and $\lambda$.  We can quantitatively analyze the change of $U_{\rm c}$
using the Mott-Hubbard criterion, 
according to which 
 a Mott transition occurs when
the atomic charge gap becomes comparable to the average kinetic
energy:
\begin{equation}
  \Delta_\text{ch}(n,U_{\rm c},J_{\rm H},\lambda) \equiv U_{\rm c}+\delta\Delta_\text{ch}(n,J_{\rm H},\lambda) = \tilde{W}(n,J_{\rm H},\lambda). 
  \label{eqn:MottHubbardCriterion}
\end{equation}
 $\Delta_\text{ch}$ is the charge gap of the local
Hamiltonian, and
$\tilde{W}(n,J_{\rm H},\lambda)$ is the average kinetic energy.
Here, $n$ is integer for commensurate Mott insulators.
Since SOC reduces the degeneracy of the atomic ground states, $\tilde{W}$ is
basically a decreasing function of $\lambda$ except for $n=3$ where
 the ground state degeneracy is not changed by introducing
SOC. By diagonalizing the local
Hamiltonian, we observe that $\delta\Delta_\text{ch}$ is an increasing
function of $\lambda$ for $n=1$, $2$, and $4$, but a decreasing
function for $n=3$ and $5$.
Altogether, for
$n=1$, $2$, and $4$, the two terms contributing to $U_{\rm c}=\tilde{W}-\delta\Delta_\text{ch}$
cooperate to reduce $U_{\rm c}$ as we also observe in our DMFT
results.  A smaller $U_{\rm c}$ at $n=4$ compared to
$n=2$  
is consistent with 
the
Mott-Hubbard criterion. In contrast, for $n=5$ the two
contributions to $U_{\rm c}$ compete and it is hard to
predict the behavior of $U_{\rm c}$ from this criterion.
  We can anticipate based on the DMFT results that
the reduction of the kinetic energy dominates the slight decrease of
the atomic gap.  Finally, at $n=3$ there is
an unchanged degeneracy and $\delta\Delta_\text{ch}$ decreases due to
 SOC implying a slight increase of $U_{\rm c}$ (compare
Fig.~\ref{fig:PhaseDiagram} (a) and (b) and see
Fig.~S4 in Ref.~\onlinecite{Aaram:2016supp}).

\begin{figure}[]
\includegraphics[width=0.45\textwidth]{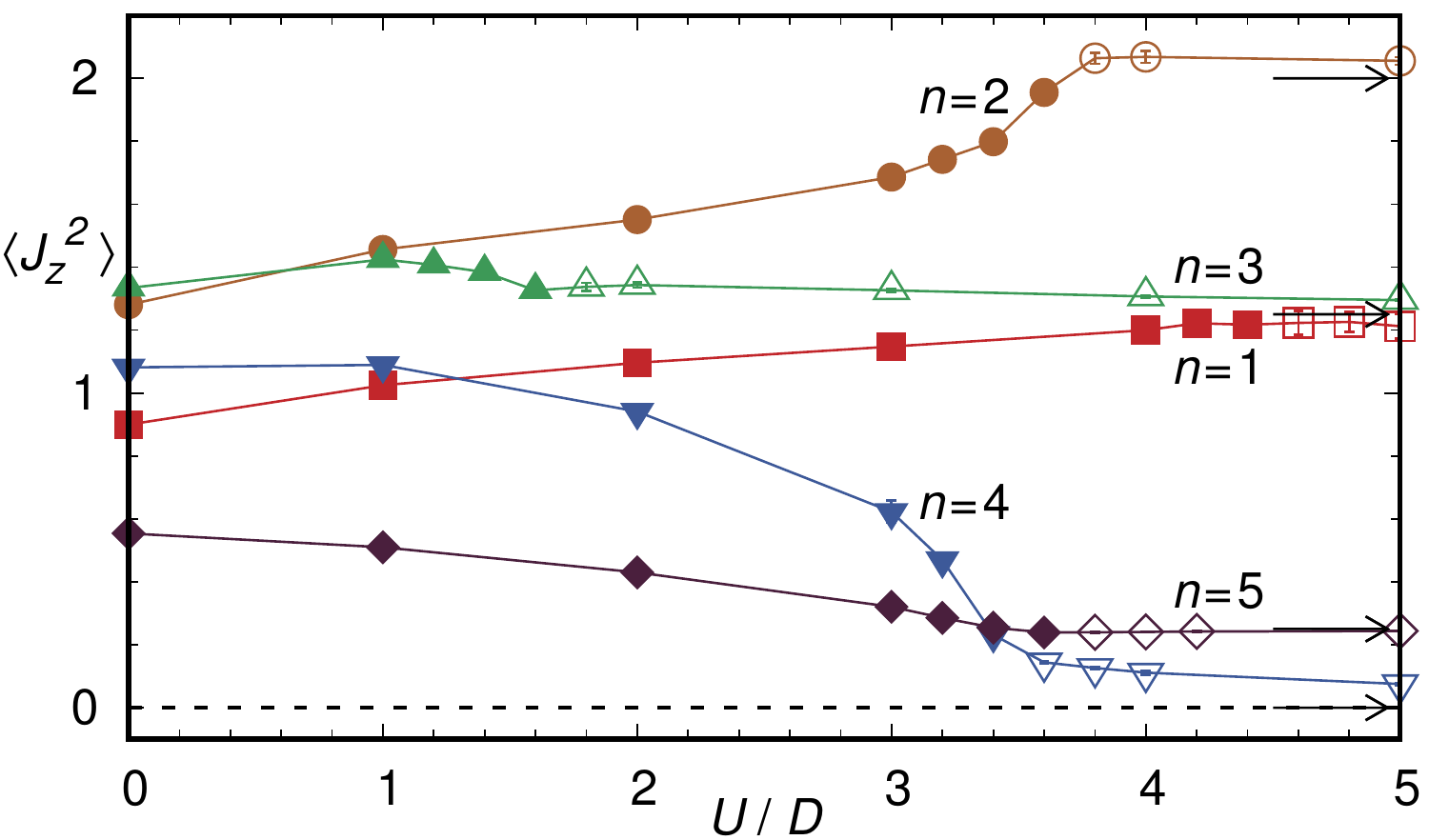}
\caption{Size of the local $J_z$-moments as a function of
    interaction strength $U/D$ for $\lambda/D=0.25$, $J_{\rm H}/U=0.15$ and
    $T/D=0.03$ at various commensurate fillings. The parameter set is
    the same as in Fig.~\ref{fig:PhaseDiagram}(b). 
    Solid (Open) symbols correspond to the metallic (insulating) solutions.
    The arrows represent the corresponding values from the Hund's rule.
  }
  \label{fig:Jzsq}
\end{figure}

The effect of the Hund's coupling  can be seen by comparing
Figs.~\ref{fig:PhaseDiagram} (b) and (c). Away from half-filling,
$U_{\rm c}$ increases with $J_{\rm H}$ but at half-filling
it slightly decreases, which is consistent with the behavior of
$\delta\Delta_\text{ch}$~\cite{deMedici:2011}.
For even stronger SOC, $\lambda/D=0.5$, a drastically reduced $U_c$ is found at $n=4$
implying an adiabatic connection of the Mott insulator to the band insulator in the 
$\lambda\gg 1$ limit~\cite{Aaram:2016supp,Du:2013}.

We now concentrate on the metallic regions.  In the spin-orbit-coupled
multiorbital system  the dynamic contribution to the susceptibility
is larger below half-filling compared to the particle-hole transformed 
state (red area in Fig.~\ref{fig:PhaseDiagram} (b) and (c))~.
Such a difference mainly comes from the cross-correlation between the spin and the orbital moment,
which is positive for $n<3$ and negative for $n>3$~(see Fig.~S3(d) in \cite{Aaram:2016supp}).
A recent study~\cite{Hoshino:2015gb}
 has shown that in the case of a multiorbital Hubbard
model without spin-orbit coupling, 
$s$-wave spin-triplet superconductivity can appear along the spin-freezing line.
The effect of the spin-orbital cross-correlation on this superconductivity 
will be an interesting future research topic.

The asymmetry in susceptibility and dynamical contribution to the susceptibility
 between below and above half-filling can be
explained by Hund's third rule whose origin is the spin-orbit
coupling~\cite{Aaram:2016supp,Bunemann:2016,Ashcroft}.  Following Hund's third rule,
in the atomic limit the alignment between $\mathbf{L}$ and
$\mathbf{S}$ depends on whether the filling is below or above
half-filling.  In our calculation, $\mathbf{L}$ and
$\mathbf{S}$ are aligned in the same direction below half-filling,
while they are anti-aligned above half-filling.
Therefore, the size of the total $\mathbf{J}$-moment is larger at
fillings below $n=3$ as we increase the interaction strength and
further localize the electrons.  Figure~\ref{fig:Jzsq} shows the
evolution of $\langle J_z^2\rangle$ as a function of Coulomb
interaction strength for five commensurate fillings and parameter
values as chosen in Fig.~\ref{fig:PhaseDiagram} (b).  In the
intermediate and strong interaction region, $U/D\gtrsim 2$, an
enhanced value of the $\mathbf{J}$-moment is found at $n=2$ and $1$
compared to the cases $n=4$ and $5$, respectively.  In the strong
correlation (Mott insulating) regime, the alignment of the
spin, orbital, and $\mathbf{J}$-moment is consistent with the atomic
results according to Hund's rules.
The $\mathbf{J}$-moment determined by atomic Hund's rule has
strong effect on the $\Delta\chi_\text{loc}$ and $\chi_\text{loc}$ 
in the metallic phase even at moderate $U$ values.

\begin{figure}[]
  \includegraphics[width=0.45\textwidth]{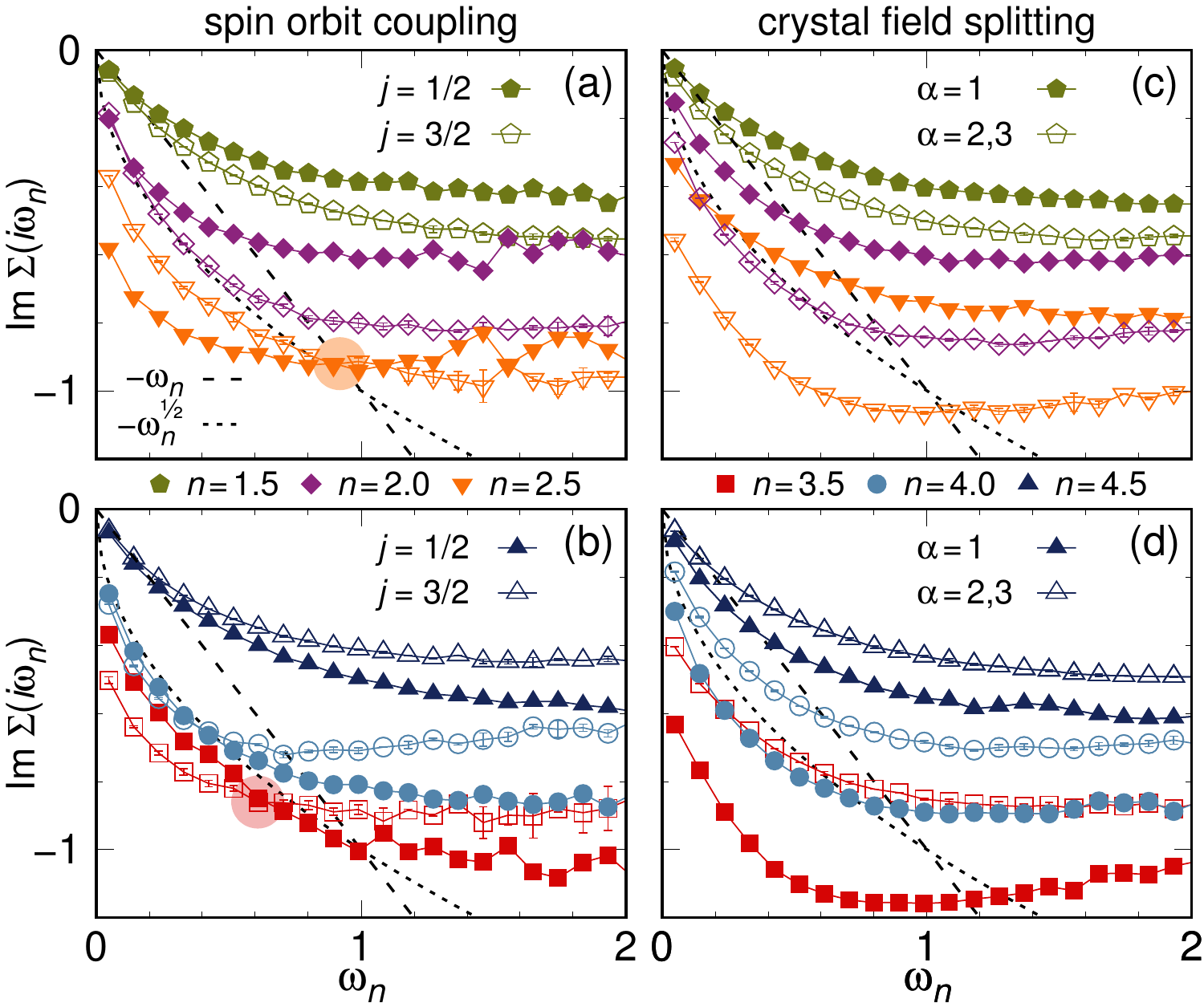}
  \caption{
    Imaginary part of the self-energy on the Matsubara axis for the system with 
    (a,b) spin-orbit coupling (SOC) (c,d) crystal-field (CF) splitting.
    For (c,d), the CF Hamiltonian, $\mathcal{H}_\text{CF}=\Delta_\text{CF}\sum^{}_{\sigma}n_{1\sigma}$ 
    is introduced instead of $\mathcal{H}_\text{SOC}$.
    The strength of the SOC and the CF are chosen to produce the same noninteracting DOS: 
    $\lambda/D=0.25$ and $\Delta_\text{CF}/D=0.375$.
    $U/D=3.0$, $J_{\rm H}/U=0.15$, and $T/D=0.015$.
    Solid (open) symbols in (a,b) denote the $j=1/2$ (3/2) results.
    Solid (open) symbols in (c,d) correspond to $\alpha=1$ ($\alpha=2,3$).
    For $j=3/2$ in (a,b) the average over $m_j=\pm 1/2,\pm 3/2$ is shown.
    In (c,d), we plot the average of $\alpha=2,3$ and spin.
    The shadings in (a,b) highlight the intersections between the different self-energies.
    The dashed (dotted) lines correspond to $-\omega_n$ ($-\omega_n^{0.5}$) 
    as a guide for the low frequency scaling.
    }
  \label{fig:ImSw}
\end{figure}

Inside the $\mathbf{J}$-freezing region
(denoted by crosses in Fig.~\ref{fig:PhaseDiagram}~(b)),
 we observe a non-Fermi liquid
(NFL) behavior of the metallic state. In order to explore this state
we show in Figs.~\ref{fig:ImSw}~(a) and (b) the imaginary part of the
self-energy on the Matsubara frequency axis across the
$\mathbf{J}$-freezing crossover line for the same parameter values
as in Fig.~\ref{fig:PhaseDiagram} (b) and various fillings.  In the low frequency region,
$\mbox{Im}\,\Sigma(i\omega_n)$ can be expressed in the form
$-\Gamma-C\omega_n^\alpha$.  As we cross the $\mathbf{J}$-freezing line, 
(region between $n\simeq 2$ and $n\simeq 4$ for $U=3$)
$\Gamma$ changes from zero to a finite value indicating a
Fermi-liquid (FL) to NFL crossover.  Near the $\mathbf{J}$-freezing
line, a small $\Gamma$ value with a non-integer exponent $\alpha$ is
found.

These two characteristic properties of the FL to NFL crossover are
reminiscent of the spin-freezing crossover observed in the model
without SOC~\cite{Werner:2008fu}.
As the system gets closer to $n=3$, the correlation function $\langle S_z(\tau)S_z(0)\rangle$ increases
while that of $\langle L_z(\tau)L_z(0)\rangle$ and $\langle L_z(\tau)S_z(0)\rangle$ decreases 
in magnitude~\cite{Aaram:2016supp}, 
so that the orbitally averaged scattering rate is determined primarily by the frozen spin moments.
However, due to SOC, the
self-energy $\mbox{Im}\,\Sigma(i\omega_n)$ of the $j=1/2$ electron is
different from that of the $j=3/2$ electrons.
  At low frequency, the
difference between $j=1/2$ and $3/2$ is enhanced in the NFL phase
compared to the FL phase.

A remarkable finding is that there exists an intersection between the
two self-energies from the different $j$ bands in the NFL phase 
(see shadings in Fig.~\ref{fig:ImSw} (a) and (b)).  This
intersection implies that the scattering rate near the Fermi-level,
$\mbox{Im}\Sigma(\omega\sim 0)$, and the total scattering rate,
$\int_{-\infty}^{\infty}d\omega~\mbox{Im}\Sigma(\omega)$ have
different relative magnitudes for the $j=1/2$ and $3/2$ electrons.
For example, for $n=3.5$, the $j=3/2$ electrons have a larger value of
$\Gamma$ with larger scattering rate at the Fermi-level, while they
exhibit a smaller high energy coefficient of the $1/(i\omega_n)$ tail,
implying a smaller total scattering rate.  Such a behavior
is not observed in the Hubbard model with ordinary crystal field (CF)
splitting (no SOC)  as shown in Figs.~\ref{fig:ImSw}~(c) and
(d)~\cite{deMedici:2009fo}.  We suggest that the basis transformation
and corresponding modification of the interaction, especially of
the Hund's coupling, are the origin of this phenomenon.  This implies
that the interplay between spin-orbit coupling effects and electronic
correlation cannot be fully captured by an effective crystal-field
splitting description. 
We call this phenomenon spin-orbit-correlation
induced flavor selectivity.

Note that the frozen $\mathbf{J}$-moment and the NFL behavior are
characteristic features of multiorbital systems with large composite moments.
Within the $\mathbf{J}$-freezing region, even the $j=1/2$ electrons show NFL
behavior, and the single-band description for $j=1/2$ is not valid
anymore.  Accordingly, the $\mathbf{J}$-freezing crossover line delimits
the region of validity of the single-band
description. 

\begin{figure}[t]
  \includegraphics[width=0.45\textwidth]{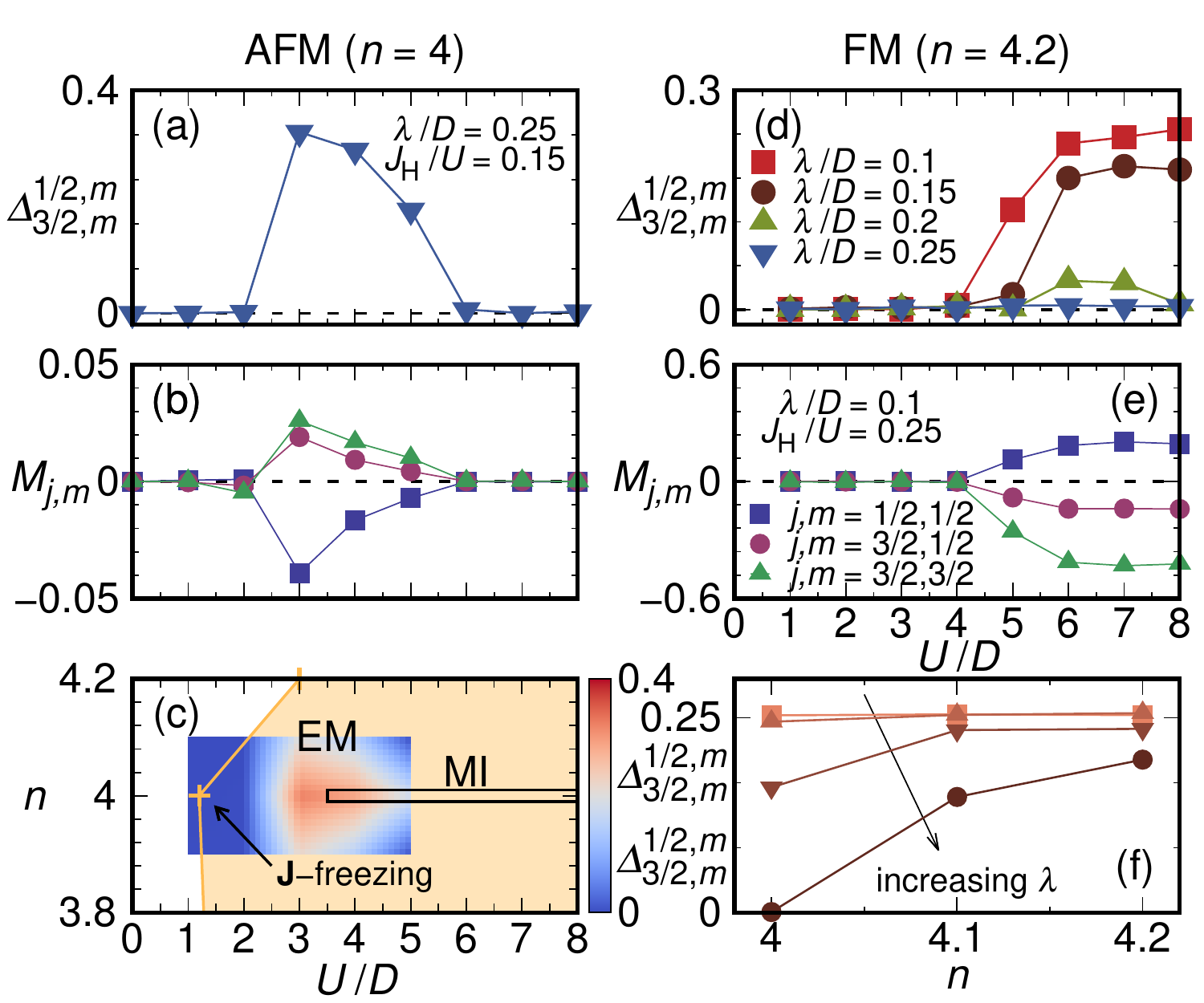}
  \caption{ (a) Excitonic order parameter and (b) magnetic
    components as a function of $U/D$ at $T/D=0.33$ for $n=4.0$,
    $\lambda/D=0.25$ and $J_{\rm H}/U=0.15$. (c) Density plot for the AFM excitonic order
    parameter.  Here, EM represents the excitonic magnetism.
    The black bar and yellow line indicate the boundary of the (paramagnetic) metal-insulator (MI) 
    and $\mathbf{J}$-freezing regime, respectively. (d) Excitonic order parameter and (e) magnetic
    components as a function of $U/D$ at $T/D=0.33$ for $n=4.2$,
     $J_{\rm H}/U=0.25$ and various  $\lambda/D$ values.
     (f) Doping dependence of the excitonic order parameter: 
     from top to bottom, corresponding $\lambda/D$ values are 0.0,0.05,0.1,and 0.15, respectively.
  }
  \label{fig:exciton}
\end{figure}

Besides the paramagnetic phase, we also investigate the
excitonic magnetism (EM) near
$n=4$~\cite{Khaliullin:2013du,Kunes:2014ea,Meetei:2015bu,Chaloupka:2016gx}.
To access such a symmetry broken phase, we introduce the off-diagonal
components of the Green function and define the order parameter of the
exciton condensed phase as $\Delta^{j'm'}_{jm}=\langle c^{\dagger}_{jm}c^{}_{j'm'}\rangle$,
where $j'\neq j$.  The magnetic components are defined as 
$M_{j,m} = \langle n_{j,+m}\rangle - \langle n_{j,-m}\rangle$.
We find two types of 
magnetism: Antiferromagnetism (AFM) and ferromagnetism (FM) at
different fillings.  At $n=4$ an AFM excitonic 
state appears at intermediate interaction
strength~\cite{Sato:2016tk,Chaloupka:2016gx,Kunes:2014ea,Hoshino:2016,Cao:2016}.
The corresponding region is located around the metal-insulator
transition point of the paramagnetic calculations, $U_{\rm c}/D\sim
3.5$.  Figures~\ref{fig:exciton}~(a) and (b) show that AFM ($M_{j,m} \neq 0$) and
excitonic order ($\Delta^{1/2,m}_{3/2,m} \neq 0$) appear simultaneously.
Upon electron doping, the AFM state is rapidly suppressed and eventually vanishes 
around $n\sim 4.2$, which is shown in Fig.~\ref{fig:exciton}(c).

Ba$_2$YIrO$_6$ is a $d^4$ system whose ground state is experimentally not completely
resolved~\cite{Dey:2016,Corredor:2016}.
According to the realistic parameter values in Ba$_2$YIrO$_6$ as given in Table~SV in the 
supplementary materials~(Ref.~\cite{Aaram:2016supp}), we would find a $J=0$ state in this system.

For large Hund's coupling and small SOC, a FM state emerges in 
the strong interaction region (Fig.~4(d) and (e)).
However, the SOC effectively suppresses the FM state and drives the 
system into an AFM state at $n=4$~(see Fig.~S5 in Ref.~\cite{Aaram:2016supp}).
Compared to the $n=4$ case, the doped FM state at $n=4.2$ is less sensitive 
to the SOC (Fig.~4(f)).
We expect that the larger kinetic energy gain for $n>4$ favors the FM state.

{\it Conclusions.}  We have explored the paramagnetic phase diagram of
the spin-orbit-coupled three-orbital Hubbard model at general filling.
We found a generalized $\mathbf{J}$-freezing crossover as a function of
$U$, $J_{\rm H}$, $\lambda$ and $n$ which exhibits a strong particle-hole
asymmetry 
and we have detected a metallic phase with a large $\Delta\chi_\text{loc}$ near
$n=2$ and a small $\Delta\chi_\text{loc}$ near $n=4$, which is the effect of Hund's third rule on the itinerant
phase.  
Across the $\mathbf{J}$-freezing line, a FL-to-NFL crossover
appears with a peculiar flavor selectivity in the NFL phase.  This is
a unique feature of SOC, which is not present in models with ordinary
crystal-field splitting. 
We expect that hole-doping of materials with $d^5$ filling like iridates
or rhodates will shift the systems toward the $\mathbf{J}$-freezing
line.
Near $n=4$, we observe excitonic magnetism
with both AFM and FM order which is consistent with a recent
mean-field study~\cite{Chaloupka:2016gx}.  Upon electron doping, the AFM state
at $n=4$ is suppressed and the FM state emerges with enhanced Hund's coupling.
These results offer new routes for finding exotic phases by doping
 $4d$ and $5d$ based materials.

{\it Acknowledgements}
We thank Ying Li, Steffen Backes, Steve Winter, Ryui Kaneko, Jan
Kune\v{s}, Alexander I. Lichtenstein, Jeroen van den Brink, Bernd B\"uchner, 
Laura Teresa Corredor Bohorquez, and Gang Cao for helpful discussions.
This research was supported by the Deutsche Forschungsgemeinschaft
through FOR1346.  The computations were performed at Center for Scientific 
Computing (CSC), the University of Frankfurt.

\clearpage


\section*{Supplementary material}

# Supplemental Information: J-freezing and Hund's rules in spin-orbit-coupled multiorbital Hubbard models

Aaram J. Kim,[1] Harald O. Jeschke,[1] Philipp Werner,[2] and Roser Valentí[1]

[1]*Institut für Theoretische Physik, Goethe-Universität Frankfurt,*
*Max-von-Laue-Straße 1, 60438 Frankfurt am Main, Germany*
[2]*Department of Physics, University of Fribourg, Chemin du Musée 3, 1700 Fribourg, Switzerland*

### $J_z$ correlation function and local susceptibility

In this section, we illustrate the correlation functions which are the basic tool for investigating the phase diagram of the three-band Hubbard model with spin-orbit coupling. Figure S1 presents the local correlation function of the total moment $J_z$ and the corresponding susceptibility for two different fillings, $n = 2$ and 4. Figure 1 in the main text and Fig. S2 are constructed based on these correlation functions. The instantaneous value of the correlation function, $\langle J_z^2 \rangle$, is determined by the size of the local moment and the right most value, $\langle J_z(\beta/2)J_z \rangle$, estimates the long-term memory for a given temperature scale. The local susceptibility $\chi_{loc}$ (Eq. (4) in the main text) and its dynamic contribution $\Delta\chi_{loc}$ (Eq. (5) in the main text) are graphically illustrated in Fig. S1 (a).

In the $n = 2$ case, the correlation function and the susceptibility show a monotonic behavior. As strong Coulomb interactions localize the electrons, both the size of the local moment and its long-term memory increase simultaneously. The increase in both quantities contributes to the susceptibility. However, the dynamic contribution shows a peak at an intermediate interaction value (Fig. S1 (c)). Compared to the noninteracting limit, the enhanced correlations and slow decay of the moment result in a larger value of $\Delta\chi_{loc}$, but this quantity again vanishes in the fully localized

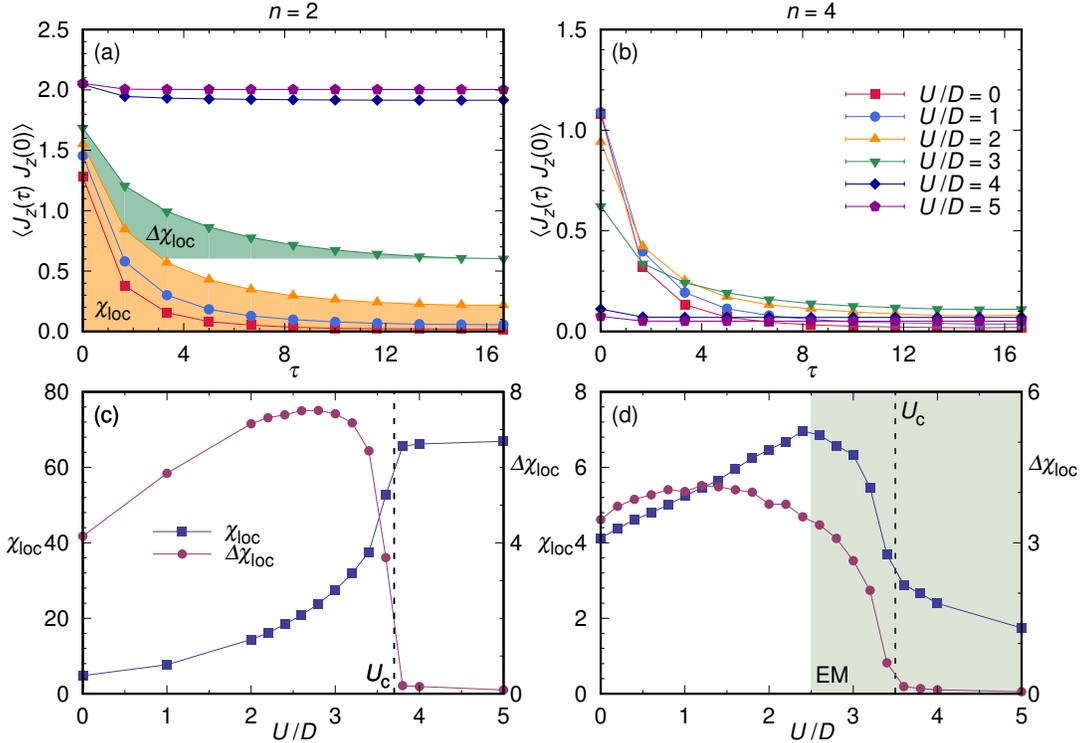

Figure S1: (a), (b) Dynamical correlation function of the total moment $J_z$ for various $U/D$ values. (c,d) The susceptibility and its dynamical contribution as a function of $U/D$. (c), (d) The vertical dashed lines mark the critical interaction strength $U_c$ of the metal-insulator-transition. The green shading in (d) represents the excitonic AFM region. The left and right columns show the $n = 2$ and $n = 4$ results, respectively. $\lambda/D = 0.25$, $J_H/U = 0.15$, and $T/D = 0.03$ for all panels. The color scheme is fixed within the same row. $\chi_{loc}$ and $\Delta\chi_{loc}$ are graphically represented in panel (a).

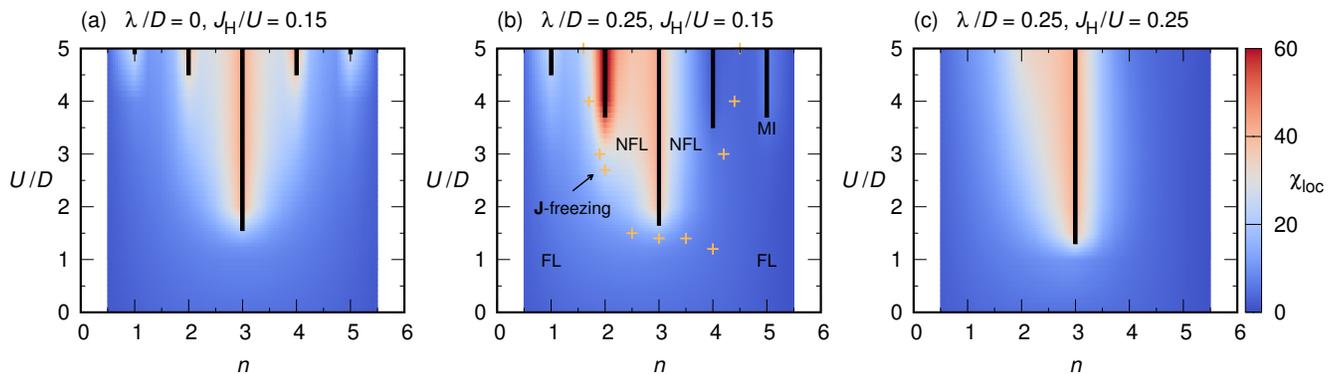

Figure S2: Local susceptibility in the $(U/D, n)$ phase diagram for (a) $\lambda/D = 0.0$, $J_H/U = 0.15$, (b) $\lambda/D = 0.25$, $J_H/U = 0.15$, (c) $\lambda/D = 0.25$, $J_H/U = 0.25$, and $T/D = 0.03$. Cross symbols mark the maximum values of $\Delta\chi_{loc}$ (compare Figs. S1 (c) and (d)) which may be used to define the **J**-freezing crossover points.

limit. Hence, a peak in $\Delta\chi_{loc}$ can be naturally expected as a function of $U$.

On the other hand, the behavior of the local susceptibility is nonmonotonic in the $n = 4$ case. The long-term memory of the correlation function and the susceptibility show a peak structure in the intermediate interaction regime. The strong suppression of the susceptibility in the localized phase, which may be attributed to the nonmagnetic character of the Van-Vleck-type ground state (vanishing **J**-moments) leads to this peak structure. The form of the ground state is shown in Table SII. In other words, the small peak in $\chi_{loc}$ is the result of a competition between **J**-freezing and quenching of the local moment. If symmetry breaking is allowed, the antiferromagnetic and excitonic order extend into the region with enhanced spin susceptibility. In this sense, we may regard the fluctuating local moments in this crossover regime as the hosting background for the symmetry-broken states.

The very different behaviors of the susceptibility in the $n = 2$ and $4$ cases are illustrated in Fig. S2. Compared to the non-spin-orbit-coupled system, Fig. S2 (a), the enhancement at $n = 2$ and the suppression at $n = 4$ of the local susceptibility in and near the localized phase are clearly evident in Fig. S2 (b). These strong effects persist into the intermediate coupling regime and for doped systems, as shown in Figs. S2 (b) and (c). The susceptibility of the itinerant phase with $n = 2$ is much larger than for $n = 4$.

It is interesting to analyze the different contributions to the total moment correlation function in the J-freezing region. The correlation function of the $J_z$-moment can be expressed as

$$\langle J_z(\tau)J_z(0)\rangle = \langle L_z(\tau)L_z(0)\rangle + \langle S_z(\tau)S_z(0)\rangle + \langle L_z(\tau)S_z(0)\rangle + \langle S_z(\tau)L_z(0)\rangle \ . \tag{1}$$

Figure S3 shows the $J_z$-correlation function and its individual terms for several different fillings. When considering fillings which are symmetric relative to $n = 3$, for example $n = 2$ and $4$ (or $n = 2.5$ and $3.5$), one notices that the difference in the $J_z$-correlations function mainly comes from the opposite sign in the $L_z$-$S_z$ contributions. On the other hand, the $L_z$- and $S_z$-correlation functions are very similar for both fillings.

As the system gets closer to half-filling, the contribution of the $S_z$-correlation function increases while that of the $L_z$ and $L_z$-$S_z$ correlation function is suppressed. The dominant contribution of the $S_z$ correlation function at $n = 2.5$ and $3.5$ can be attributed to the effect of the atomic ground state at $n = 3$ with a spin-maximized and orbital-singlet state. This analysis suggests that the flavor-averaged value of the scattering rate at the Fermi-level is mostly determined by the spin correlation function and that the similar values of the dominant $S_z$-correlation function for symmetric fillings result in similar values of the average scattering rate. This is indeed the case as can be seen in Fig. 3 of the main text.

### Local Hamiltonian

The local Hamiltonian not only describes the physical properties of the fully localized limit but also affects the properties of the correlated itinerant phases. In our model, the local Hamiltonian is composed of two terms: the SOC and the Kanamori-type Coulomb interaction,

$$\mathcal{H}_{loc} = \mathcal{H}_\lambda + \mathcal{H}_{int} \ . \tag{2}$$

Using the total electron number $N$ and the total moment $J_z$ as quantum numbers, we diagonalized $\mathcal{H}_{loc}$. Tables SI and SII summarize the properties of the ground states for each $N$. In Table SI, the numbers in parenthesis are the

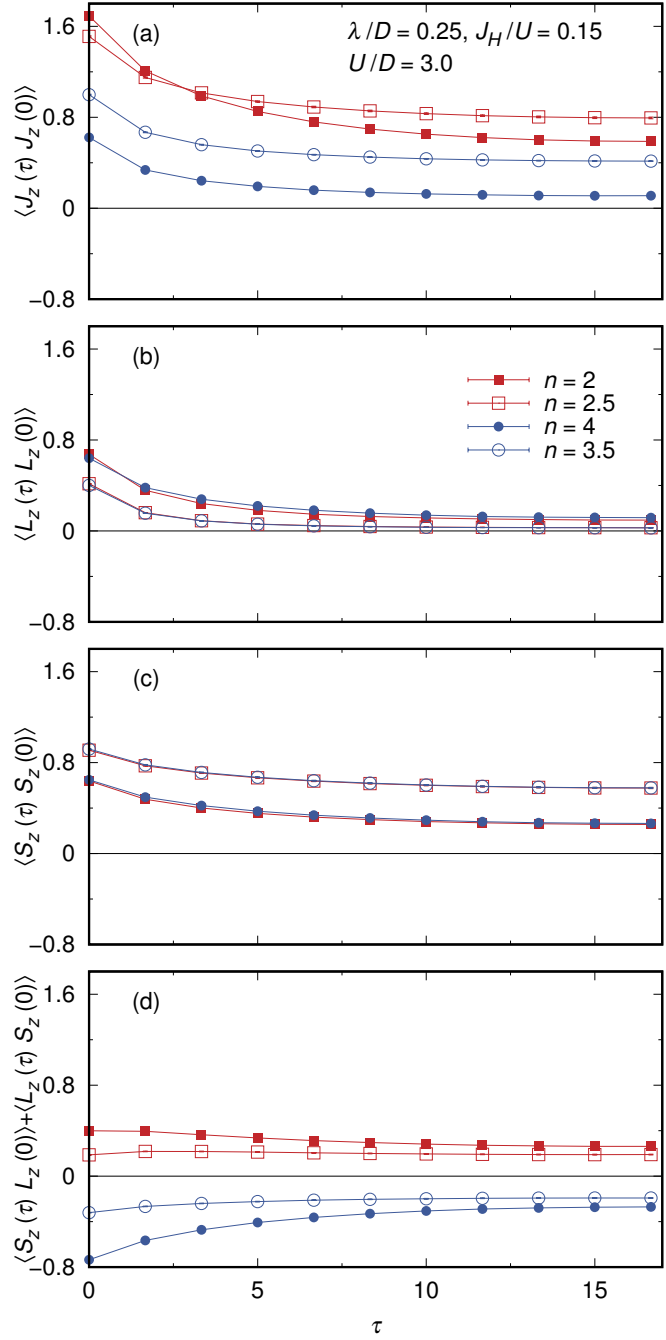

Figure S3: Various correlation functions for $\lambda/D = 0.25$, $J_{\rm H}/U = 0.15$, $U/D = 3$, and $T/D = 0.03$. From top to bottom, the corresponding functions are the auto-correlation functions of the $J_z$, $L_z$, and $S_z$ moment, and the cross-correlation function between the $S_z$ and $L_z$ moment.

ground state degeneracies of the non-spin-orbit-coupled Hamiltonian and $\mu$ is the chemical potential. As we turn on the SOC, the degeneracy is reduced except for $N = 3$, which means that the average kinetic energy is reduced, as discussed in the main text. Based on the ground state energy, we calculate the charge gap, defined by

$$\Delta_{\rm ch}(N) = (E_{\rm g}(N+1) - E_{\rm g}(N)) - (E_{\rm g}(N) - E_{\rm g}(N-1)) \ , \tag{3}$$

where $E_{\rm g}(N)$ is the ground state energy of the local Hamiltonian for filling $N$. The results are summarized in Table SIII, and Fig. S4 represents $\delta\Delta_{\rm ch}(N, J_{\rm H}, \lambda) \equiv \Delta_{\rm ch} - U$. The summary of the Hund's rules which apply to the local Hamiltonian is also presented in Table SIV.

| $N$ | Degeneracy | Ground State Energy ($E_g$) |
|---|---|---|
| 0 | 1 (1) | 0 |
| 1 | 4 (6) | $-\frac{\lambda}{2} - \mu$ |
| 2 | 5 (9) | $\frac{1}{4}\left(4U - 8J_H - \lambda - \sqrt{16J_H^2 + 8J_H\lambda + 9\lambda^2}\right) - 2\mu$ |
| 3 | 4 (4) | $3U + f(J_H, \lambda) - 3\mu$ |
| 4 | 1 (9) | $\frac{1}{2}\left(12U - 21J_H - \lambda - \sqrt{25J_H^2 + 10J_H\lambda + 9\lambda^2}\right) - 4\mu$ |
| 5 | 2 (6) | $10U - 20J_H - \lambda - 5\mu$ |
| 6 | 1 (1) | $15U - 30J_H - 6\mu$ |

Table SI: Ground state energy and degeneracy of the local Hamiltonian for different total particle numbers. The numbers in the parentheses show the ground state degeneracies when $\lambda = 0$.

Table SII: Ground state for a given sector of the local Hamitonian. In our notation, the upper (lower) level represents the $j = 1/2$ (3/2) states and the lower left (right) level corresponds to $m_j = \pm 1/2$ ($m_j = \pm 3/2$). Full (empty) circles mark the positive (negative) $m_j$ electron. All coefficients can be chosen to be real and duplicated symbols in different sectors have nothing to do with each other.

Figures S4 (b) through (f) illustrate the effect of the SOC on the charge gap, which turns out to be different for different $N$. In the regime of interest, finite $J_H$ and relatively small $\lambda$, $\delta\Delta_{ch}$ is an increasing function of $\lambda$ for $N = 1$, 2 and 4, but a decreasing function for $N = 3$ and 5. Based on the information of the degeneracy and charge gap, we discuss the qualitative behavior of the critical interaction strength $U_c$ in the main text.

| $N$ | Charge Gap ($\Delta_{\mathrm{ch}}$) |
|---|---|
| 1 | $U - 2J_{\mathrm{H}} + \frac{3\lambda}{4} - \frac{1}{4}\sqrt{16J_{\mathrm{H}}^2 + 8J_{\mathrm{H}}\lambda + 9\lambda^2}$ |
| 2 | $f(J_{\mathrm{H}},\lambda) + U + 4J_{\mathrm{H}} + \frac{1}{2}\sqrt{16J_{\mathrm{H}}^2 + 8J_{\mathrm{H}}\lambda + 9\lambda^2}$ |
| 3 | $U - \frac{25J_{\mathrm{H}}}{2} - \frac{3\lambda}{4} - \frac{1}{4}\sqrt{16J_{\mathrm{H}}^2 + 8J_{\mathrm{H}}\lambda + 9\lambda^2} - \frac{1}{2}\sqrt{25J_{\mathrm{H}}^2 + 10J_{\mathrm{H}}\lambda + 9\lambda^2} - 2f(J_{\mathrm{H}},\lambda)$ |
| 4 | $f(J_{\mathrm{H}},\lambda) + U + J_{\mathrm{H}} + \sqrt{25J_{\mathrm{H}}^2 + 10J_{\mathrm{H}}\lambda + 9\lambda^2}$ |
| 5 | $U - \frac{J_{\mathrm{H}}}{2} - \frac{\lambda}{2} - \frac{1}{2}\sqrt{25J_{\mathrm{H}}^2 + 10J_{\mathrm{H}}\lambda + 9\lambda^2}$ |

Table SIII: Charge gap for different total particle numbers. Note that $\Delta_{\mathrm{ch}}$ can be expressed as $U + \delta\Delta_{\mathrm{ch}}(N, J_{\mathrm{H}}, \lambda)$. We do not specify $f(J_{\mathrm{H}}, \lambda)$ because it is difficult to express it in a simple form. Numerical values are used to obtain Fig. S4.

| $n$ | $S$ | $L$ | $J$ |
|---|---|---|---|
| 1 | 1/2 | 1 | 3/2 |
| 2 | 1 | 1 | 2 |
| 3 | 3/2 | 0 | 3/2 |
| 4 | 1 | 1 | 0 |
| 5 | 1/2 | 1 | 1/2 |

Table SIV: Summary of Hund's rules including the third law.

### Energy Scales in Various Materials

In Tab. SV, we summarize the previous estimates of the energy scales for various materials. Our main parameter set is chosen to be the estimate from $Sr_2IrO_4$ [1]. The absolute value of the SOC is mainly determined by the relevant atom; 0.1∼0.2 eV for Ru, Rh and 0.3∼0.4 eV for Os, Ir. In these materials, the Hund's coupling strength $J_H/U$ varies within 0.1∼0.22 which is the range considered in our calculations.

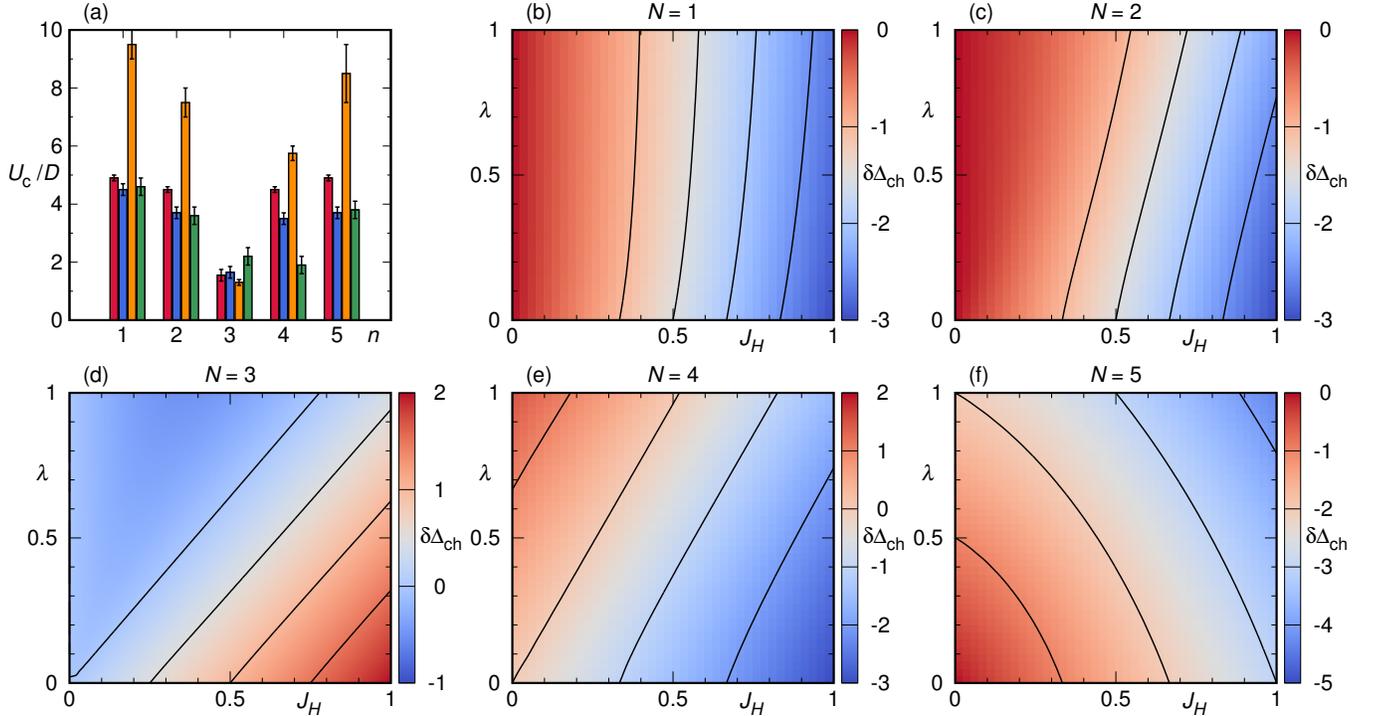

Figure S4: (a) Critical interaction strength $U_{\mathrm{c}}$ as a function of electron filling. From left to right within a group of four bars, the bars represents the $U_{\mathrm{c}}$ for $(\lambda, J_H/U) = (0, 0.15)$, $(0.25, 0.15)$, $(0.25, 0.25)$, and $(0.5, 0.15)$, respectively. (b)-(f) Density plot of $\delta\Delta_{\mathrm{ch}}$ as a function of $J_H$ and $\lambda$ for different total particle number $N$. $\delta\Delta_{\mathrm{ch}}(N, J_{\mathrm{H}}, \lambda)$ is defined as $\Delta_{\mathrm{ch}} - U$.

| Material | Conf. | $D$ | $\lambda/D$ | $J_H/U$ | $U/D$ | Ref. |
|---|---|---|---|---|---|---|
| $Sr_2IrO_4$ | $d^5$ | 1.44eV | 0.257 (0.37eV) | 0.15 | - | [1] |
| $Sr_2RhO_4$ | $d^5$ | 1.44eV | 0.125 (0.18eV) | 0.15 | - | [1] |
| $Y_2Ir_2O_7$ | $d^5$ | 0.5eV | 0.8 (0.4eV) | 0.1 | 5.0 (2.5eV) | [2] |
| $Na_2IrO_3$ $Li_2IrO_3$ | $d^5$ | 0.2eV | 2.0 (0.4eV) | 0.2 | 15 (3eV) | [3] |
| $\alpha$-$RuCl_3$ | | | 0.75 (0.15eV) | | | |
| $Sr_2RuO_4$ | $d^4$ | 0.75eV | 0.13 (0.100eV) | - | - | [4] |
| $Sr_2YRuO_6$ | $d^4$ | 0.55eV | 0.18 (0.100eV) | 0.20 | 4.73 (2.6eV) | [5–7] |
| $Sr_2YIrO_6$ $Ba_2YIrO_6$ | $d^4$ | 0.5eV | 0.6 (0.33eV) | 0.22 | 3.6 (1.8eV) | [8] |
| $Sr_2CrOsO_6$ | $d^3$ | 0.75eV | 0.4 (0.3eV) | 0.17 | 2.67 (2eV) | [9] |
| $Ba_2NaOsO_6$ | $d^1$ | 0.3eV | 1.0 (0.3eV) | - | - | [10, 11] |

Table SV: Summary of the energy scales in various spin-orbit-coupled materials. $D, \lambda, J_H$, and $U$ are the half-bandwidth, strength of spin-orbit coupling, Hund's coupling, and Coulomb interaction, respectively.

### Long-Range Order near $n = 4$

For large Hund's coupling and small SOC, we observe itinerant FM. In Fig. S5, the increase of the SOC results in successive phase transitions from FM to AFM, and from AFM to PM. For a given total moment $\langle \mathbf{J} \rangle$, the internal structure is different depending on the magnetic phase. In the FM phase, $\langle \mathbf{S} \rangle$ and $\langle \mathbf{L} \rangle$ are aligned in the same direction, and $\langle \mathbf{S} \rangle$ makes the dominant contribution to $\langle \mathbf{J} \rangle$. On the other hand, in the AFM phase, $\langle \mathbf{S} \rangle$ and $\langle \mathbf{L} \rangle$ are anti-aligned, but $\langle \mathbf{L} \rangle$ dominates $\langle \mathbf{S} \rangle$.

Our results show that the Hund's coupling enhances the FM while the SOC suppresses it. This behavior is consistent with the previous results [5, 12], especially two-site Exact diagonalization (ED) results in Fig. S6. Assuming a square lattice, the value of the critical strength of the SOC, $\lambda_c/D = 0.1275$ for $J_H/D = 0.25$ in the ED results is compatible with our DMFT results.

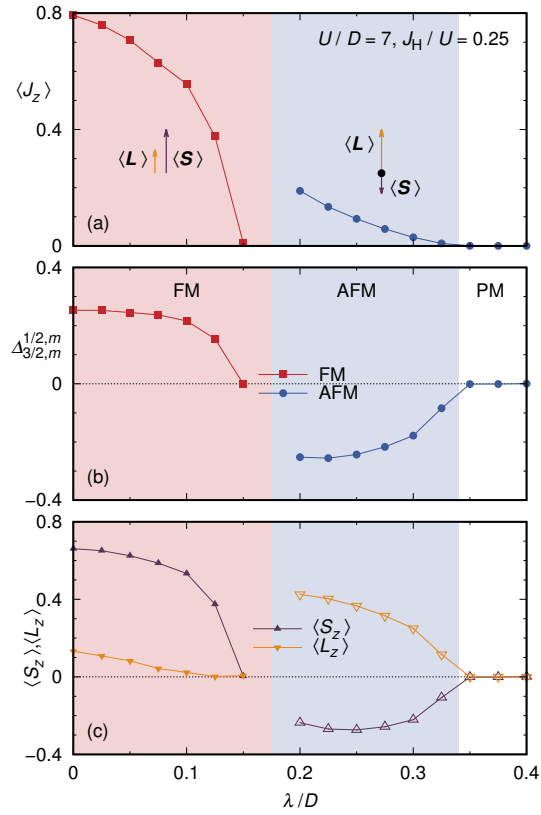

Figure S5: (a) Total magnetic moment, (b) excitonic order parameter and (c) spin/orbital moment as a function of $\lambda/D$ for $T/D = 0.33$, $n = 4$, $J_H/U = 0.25$, and $U/D = 7$. The red (blue) shading represents the FM (AFM) region. The arrows in (a) show the alignment (anti-alignment) of spin and orbital moment in the FM (AFM) region.

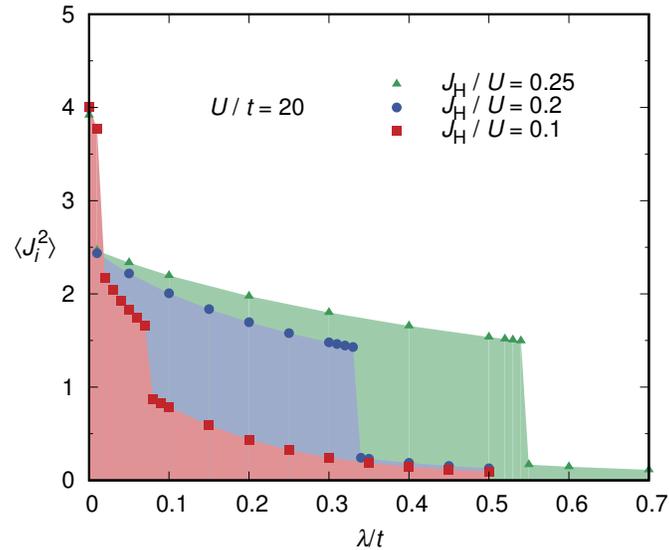

Figure S6: The size of the total moment for a given site in the two-site Hubbard model as a function of $\lambda/t$ for $U/t = 20$. Here $t$ is the hopping parameter between two sites. From top to bottom, the corresponding Hund's coupling strengths are 0.25, 0.2, and 0.1, respectively.



\end{document}